\documentclass[10pt,times,twocolumn]{article}

\usepackage{epsfig,mathptmx,times,amsmath,color,url}
\usepackage{bm,amsmath}
\usepackage{graphicx}
\usepackage{caption}
\usepackage{subcaption}
\graphicspath{Figures/}
\usepackage{algorithmic}
\usepackage{algorithm}
\usepackage{graphicx}
\graphicspath{Figures/}

\newcommand{\pv}{\mathbf{p}}

\newcommand{\xv}{\mathbf{x}}

\newcommand{\Gr}{\mathcal{G}}
\newcommand{\G}{{\mathbf{g}}}
\newcommand{\Go}{{\mathbf{go}}}
\newcommand{\Lo}{\mathbf{l}}

\newcommand{\R}{\mathbf{r}}

\newcommand{\V}{\mathcal{V}}
\newcommand{\E}{\mathcal{E}}

\newcommand{\scs}{\chi}
\newcommand{\Scs}{\Xi}

\newcommand{\tstep}{\delta}
\newcommand{\Neb}{\mathrm{Neb}}

\newcommand{\pvnot}{\pv^0}

\newcommand{\Pen}{\mathrm{Pen}}
\newcommand{\param}{\alpha}
\newcommand{\paramp}{\alpha^P}
\newcommand{\parami}{\alpha^I}
\newcommand{\paramf}{\alpha^F}
\newcommand{\paraml}{\beta^l}
\newcommand{\gi}{g \to i}
\newcommand{\ije}{i \to j}
\newcommand{\pvg}{\pv^{\Go}}
\newcommand{\pvgnot}{\pv^{\Go}_{0}}
\newcommand{\pvl}{\pv^{\Lo}}
\newcommand{\pvlnot}{\pv^{\Lo}_{0}}
\newcommand{\pvr}{\pv^{\R}}
\newcommand{\xvd}{\mathbf{s}^{d}}
\newcommand{\angl}{\theta}
\newcommand{\disc}{\gamma}
\newcommand{\GenCost}{\mathrm{GenCost}}
\newcommand{\pvgl}{\underline{\pvg}}
\newcommand{\pvgh}{\overline{\pvg}}
\newcommand{\pvgrl}{\underline{\pvg_r}}
\newcommand{\pvgrh}{\overline{\pvg_r}}
\newcommand{\pvgi}{\pv^{I}}
\newcommand{\pvh}{\overline{\pv}}
\newcommand{\Ilim}{\overline{E^{\Go}}}

\begin{document}
\title{Distributed Control of Generation in a Transmission Grid with a High Penetration of Renewables}

\author{Krishnamurthy~Dvijotham
Dept. of Computer Science and Engineering\\
University of Washington\\
Seattle, WA, 98195\\
Email: dvij at cs.washington.edu \\
Scott Backhaus
MPA Division\
Los Alamos National Laboratory\\
Los Alamos, NM 87545\\
Email: backhaus at lanl.gov
Michael Chertkov 
Theory Division \& CNLS \\
Los Alamos National Laboratory\\
Los Alamos, NM 87545\\
Email: chertkov at lanl.gov
}

\maketitle

\begin{abstract}
Deviations of grid frequency from the nominal frequency are an indicator of the global imbalance between generation and load.  Two types of control, a distributed proportional control and a centralized integral control, are currently used to keep frequency deviations small. Although generation-load imbalance can be very localized, both controls primarily rely on frequency deviation as their input.  The time scales of control require the outputs of the centralized integral control to be communicated to distant generators every few seconds.  We reconsider this control/communication architecture and suggest a hybrid approach that utilizes parameterized feedback policies that can be implemented in a fully distributed manner because the inputs to these policies are local observables at each generator.  Using an ensemble of forecasts of load and time-intermittent generation representative of possible future scenarios, we perform a centralized off-line stochastic optimization to select the generator-specific feedback parameters.  These parameters need only be communicated to generators once per control period (60 minutes in our simulations). We show that inclusion of local power flows as feedback inputs is crucial and reduces frequency deviations by a factor of ten.  We demonstrate our control on a detailed transmission model of the Bonneville Power Administration (BPA). Our findings suggest that a smart automatic and distributed control, relying on advanced off-line and system-wide computations communicated to controlled generators infrequently, may be a viable control and communication architecture solution. This architecture is suitable for a future situation when generation-load imbalances are expected to grow because of increased penetration of time-intermittent generation.

\end{abstract}

\maketitle

\section{Problem Setup and Brief Statement of Results}

In today's power systems, the system operator performs an Optimal Power Flow (OPF) dispatch periodically with typical time interval being 5, 15, or 60 minutes depending on the Balancing Area \cite{Kundur}.  The OPF sets the power outputs of the committed generation to match power demand and minimize generation cost while respecting the capacity limits on lines, ramping constraints and limits on generators and sometimes taking into account the N-1 security constraints. In between two successive OPFs, the system is automatically controlled by a combination of two mechanisms.  The faster of the two, acting on the scale of seconds, is primary frequency control--a fully distributed proportional feedback on locally-measured frequency deviations that may also include a deadband.  The slower mechanism, acting on the scale of minutes, is automatic generation control (AGC), also called secondary control--a centralized feedback on the integral of a weighted sum of a centrally measured frequency and tie line flows to neighboring balancing areas.\cite{KT-DEB-VV-BA:05}\\

These combined controls correct deviations in the generation-load balance driven by fluctuations in loads, renewables and other disturbances in the system. However, these mechanisms do not explicitly incorporate line-flow limits, generators ramping limits, or time-integral constraints like those on run-of-river hydro generation or energy storage. For systems with relatively low levels of fluctuations, these limits are not frequently violated and it is not necessary incorporate them directly. However, higher levels of time-intermittent generation will create larger fluctuations and ramping events and the associated constraint violations will become more common.  Standard primary and secondary controls are limited in their ability to balance these fluctuations, and better control design is needed to manage these larger fluctuations.  Because these fluctuations are intimately connected to frequency deviations, they are of special concern because they my result in  system-wide instabilities and loss of synchrony \cite{FERC-frequency}. \\

Other considerations for real-time power grid control systems are communication constraints and communication security\cite{KT-DEB-VV-BA:05}. Mechanisms that rely on central aggregation of the entire grid state followed by a centrally computed response will be vulnerable to communication failures and attacks on the communication network, making the overall system less robust. On the other hand, with significant renewable penetration, it is difficult to control a system purely based on local feedback, since under some conditions, it may be necessary to control distant generators in a correlated manner.\\

In this preliminary work, we explore a hybrid approach that combines the speed and security of fully distributed control with the extensive system visibility provided by centralized control.  Our method performs a centralized lookahead dispatch that also computes optimal local feedback parameters for all controllable generation, thus enabling the system to respond to fluctuations based only on local observables.  We expand our definition of local observables to include not just frequency but also real power flows to neighboring nodes.  We use an ensemble of forecasts that capture various possible scenarios for the wind generation and loads over the next intra-dispatch period (5 min/15 min/ 1 hour) to design an optimal time-varying dispatch for all the generators, as well as local feedback functions that enable the generators to respond to fluctuations based on the local observables.\\

\begin{figure}
\begin{center}
\includegraphics[width=.3\textwidth]{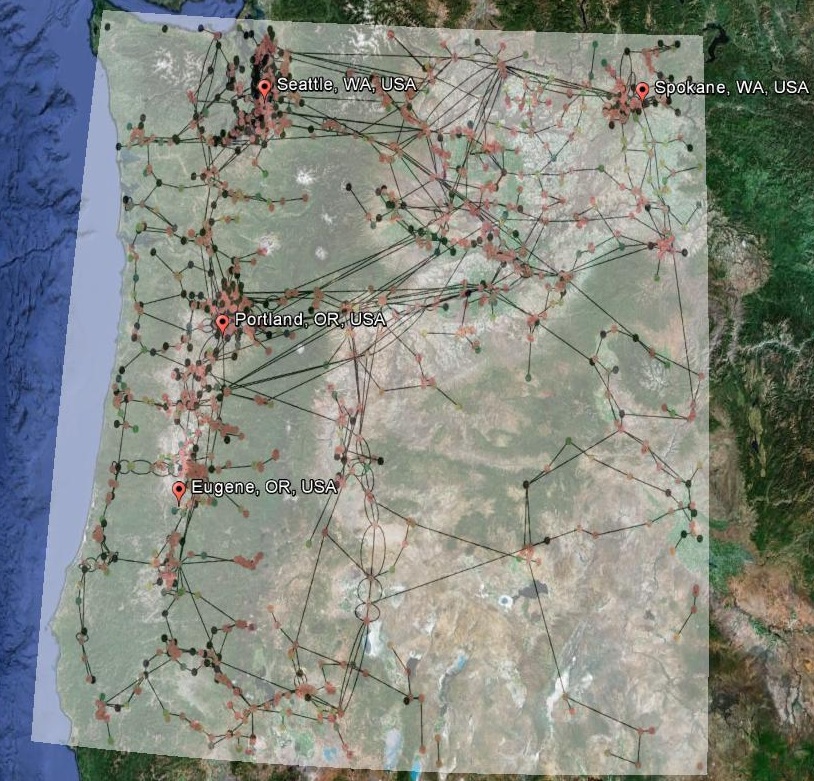}
\caption{Our model of the BPA Transmission Network} \label{fig:BPA}
\end{center}
\end{figure}

Our control design is split into 2 phases:
\begin{itemize}
\item[a] An off-line optimization phase where the distributed control gains are optimized jointly for the whole network in a central computer using extensive simulation of possible future wind generation and forecast scenarios. These gains are then communicated to each flexible resource (controllable device) in the transmission network. This off-line optimization would need to be re-run every time the statistics of possible future scenarios change significantly. In general, we expect this optimization to be run every time the generation re-dispatch changes.
\item[b]  An online response phase where each device implements its purely local control in response to local observables (local frequency, line flows etc.) on the pace of the standard primary controls.
\end{itemize}
We test our algorithm on historical data from the Bonneville Power Administration (BPA) system \cite{BPA}, an ideal test system for our algorithm as it has significant amounts of both hydro and wind generation.  We show that our algorithm performs well, even in cases of significant wind ramps.

Our results (detailed in section \ref{sec:Numerics}) lead to the following important observations:
\begin{itemize}
\item[a] Local control based on response to frequency deviations and local line flows at each generation can keep frequency deviations down to the about 10 mHz while maintaining all the security and capacity constraints.
 \item[b] Proportional control on frequency deviations and feedback on line flows is sufficient.  Adding a frequency-deviation integral response is unnecessary, which is advantageous because a distributed implementation of an integral term may cause instabilities due to errors in local frequency measurement, and also because it limits communication requirements.
  \item[c] Joint optimization of feedback parameters for frequency deviation and line flows is necessary. Independent optimization or removal of either term leads to poor control performance.
   \item[d] Optimization over a finite but representative set of future scenarios enables the generalization of the control to new unseen scenarios.
\end{itemize}

The rest of the paper is organized as follows: Section \ref{sec:Math} describes the mathematical setting of the underlying control/optimization problem; we describe and discuss results of our numerical BPA experiments in Section \ref{sec:Numerics}; and Section \ref{sec:Future} presents conclusions and explains our path forward.

\section{Mathematical Formulation}
\label{sec:Math}

\subsection{Preliminaries}

The power system is described by an undirected graph $\Gr=(\E,\V)$ with edges $\E$ and $n$ vertices $\V$. The grid is composed of loads ($\Lo$), conventional generators ($\G$) and renewable generators ($\R$). The flexible resources in our grid are the online conventional generators $\Go$. We denote by $\pv$ the $|\V| \times 1$ vector of net active power injections at each node in the network and by $\pvg,\pvr,\pvl$ the net active power injections due to online conventional generators, renewable generators and loads: $\pv=\pvg+\pvr+\pvl$. Each of these vectors is of size $|\V|\times 1$ with the convention that $\pvg_{i}=0$ or $\pvr_{i}=0$ if there is no conventional or renewable generator at node $i$. Note here that we make the assumption that there is only one generator or load at a given node. If there are multiple, we replace them by an equivalent single generator or load.  For a vector $v$ with indices in $\V$, $v_{i}$ denotes a particular component for $i \in \V$ and $v_{S}$ denotes the sub-vector $\{v_i : i \in S\}$ for a subset $S \subset \V$.

\subsection{Scenarios}

The control algorithm proceeds by analyzing an ensemble possible future scenarios and designs control strategies that optimize the system cost (defined below) across all scenarios. We define a scenario $\scs$ to be a collection of the following quantities:
\begin{itemize}
\item[a] Renewable generation over the time horizon of interest: $\pvr(t)$.
\item[b] Load profile over the time horizon of interest: $\pvlnot(t)$.
\item[c] A unit commitment (configuration of generators which are online,  i.e. available for re-dispatch) $\Go$.
\end{itemize}
To define the control problem, we require a collection of scenarios $\Scs$ and estimates for the probability of each scenario, i.e. $\Scs=\{\scs_i,\mathrm{Prob}(\scs_i)\}$. We note that for a given collection $\Scs$, $\pvr(t)$ and $\pvlnot(t)$ (items a and b from above) will vary across the ensemble of scenarios, however, we take $\Go$ (the unit commitment from c) fixed because we are designing the time-dependent dispatch and local feedback parameter for that particular $\Go$.  In this work, we assume that the collection $\Scs$ is finite. Typically, $\Scs$ will be built up from load and wind forecasts from different forecasting methodologies weighted by confidences in each of these forecasts. $\Scs$ could also include samples from a stochastic forecasting model based on climate models, historical data, meteorological sensors etc.

%
%

\subsection{Control Formulation} \label{sec:Dyn}

We ignore electro-mechanical dynamical transients and work with a discrete-time quasi-static approximation of the system dynamics with fixed time step $\tstep$ and integer time indices $t=0,1,\ldots,T$: at each time step the power flows over lines are re-computed for configuration of consumption/generation at nodes evolving in discrete time. In general, the feedback can depend on any of the system variables, but we limit ourselves to local observables so that the control can be implemented in a completely distributed fashion at each generator after the dispatch and feedback parameters have been communicated.\\

For each generator $g \in \Go$, we compute a time-varying dispatch $\pvnot_{g}(t):0 \leq t \leq T$,  proportional frequency response coefficient $\paramp_{g}$, integral frequency response coefficient $\parami_{g}$, and a response coefficient to local flows $\{\paramf_{\gi}: i \in \Neb(g)\}$. Further, we denote by $\omega(t)$ the frequency deviation from the nominal frequency ($50/60$ Hz) at time $t$ and by $\Omega(t)$ the integral of the frequency deviation, which in discrete-time is approximated by $\Omega(t)=\sum_{\tau=0}^{t} \disc^{\tau-t}\omega(\tau)$ where $0<\disc<1$ is a discount factor. In other words, the integral frequency term is simply a weighted sum of frequency deviations in the past, where frequency deviations that are further in the past receive a geometrically smaller weight. With the time varying dispatch and feedback parameters determined, the output of the generators is given by:
\begin{align*}
\pvg_{g}(t)& ={\pvgnot}_{g}(t)+\paramp_{g}\omega(t)+\parami_{g}\Omega(t)+\\
& \quad \sum_{i \in \Neb(g)} \paramf_{\gi}\pv_{\gi}(t).
\end{align*}
Although our algorithm can incorporate nonlinear feedback, we choose feedback which is linear in the local observables for this initial work.  In addition to generators, the real power consumption of loads responds to frequency changes, and we assume a simple linear load-frequency response given by
\[\pvl(t)=\pvlnot(t)+\paraml\omega(t),\]
where the $\paraml$ are known from measurement where $\pvlnot$ is the load at the nominal frequency (60 Hz).  Combining the load and generator frequency response and the generators' time varying dispatch, the system's equilibrium frequency is computed by enforcing power balance in the system:
\begin{align*}
\sum_{i \in \V} \pvg_{i}(t)+\pvl_{i}(t)+\pvr_{i}(t)=0 \implies \nonumber \\
 \omega(t) = -\frac{\sum_i {\pvlnot}_i(t)+\pvr_i(t)+\pvg_i(t)}{\sum_i \paraml_i}.
 \end{align*}\\
To compute power flow from the injections $\pv(t)=\pvl(t)+\pvr(t)+\pvg(t)$, we use a modified version of the DC Power flow equations based on a linearization of the AC Power flow equations around the nominal dispatch at the beginning of the control period $\pv(0)$. The linearization gives us dynamic impedances $\xvd_{\ije}$ that substitute for the line reactances in the DC power flow equations:
\begin{align*}
\pv_i(t) = \pvnot+\sum_{j \in \Neb(i)} \frac{\angl_{i}(t)-\angl_{j}(t)}{\xvd_{\ije}},\nonumber\\
\pv_{\ije}(t)={\pvnot}_{\ije}+\frac{\angl_{i}(t)-\angl_{j}(t)}{\xvd_{\ije}}.
\end{align*}
Such a linearization is reasonable assuming that the flow patterns do not change too much during the course of the control period.\\

In addition to several other constraints discussed below, we will also imposed a constraint on the total energy extracted from generators in the control period.  Such constraints can represent the water discharge constraints on run-of-river hydro systems or state-of-charge constraints on energy storage devices.  Therefore, we must also include the total energy extracted from each generator into the system state:
\[\pvgi(t)=\sum_{\tau=0}^{t} \pvg(\tau).\]\\

%
%
The overall system state consists of $\xv(t)=[\Omega(t);\pvg(t);\pvgi(t)]$  (in Matlab notation), and the system evolution can be summarized by:
\begin{align}
 \omega(t) & = -\frac{\sum_i {\pvlnot}_i(t)+\pvr_i(t)+\pvg_i(t)}{\sum_i \paraml_i} \label{eq:Dyn} \\
 \Omega(t+1) & = \omega(t)+\disc \Omega(t) \nonumber \\
 {\pvg}_{g}(t) & ={\pvgnot}_{g}(t)+\paramp_{g}\omega(t)+\parami_{g}\Omega(t) \nonumber \\
 & \quad +\sum_{i \in \Neb(g)} \paramf_{\gi}\pv_{\gi}(t) \nonumber  \\
 \pvl(t) & =\pvlnot(t)+\paraml\omega(t) \nonumber \\
\pv_i(t) & = \sum_{j \in \Neb(i)} \frac{\angl_{i}(t)-\angl_{j}(t)}{\xvd_{\ije}},\pv_{\ije}(t)=\frac{\angl_{i}(t)-\angl_{j}(t)}{\xvd_{\ije}} \nonumber
\end{align}
%
\subsection{Cost Functions}

We consider a stochastic setting with many possible features, and it is unclear whether it is feasible to satisfy all constraints across all scenarios in $\Scs$. Therefore, we use a penalty function to enforce our constraints in a smooth manner.  The penalty function has a magnitude of zero in a dead-band around the most feasible region and grows cubically with the magnitude of constraint violation:
\[\Pen(a,l,u)= \begin{cases}
10^7{\left((a-u)/(0.1*(u+1))\right)}^3 & \text{ if } a \geq u \\
10^7{\left((l-a)/(0.1*(l+1))\right)}^3 & \text{ if } a \leq l \\
0 & \text{ otherwise }
\end{cases}.\]
Here, $a$ is the value of the constrained quantity and $l$ and $u$ are the lower and upper bounds on $a$, respectively. We also adopt the convention that when $a,l,u$ can be vectors (of the same size) and the penalty in this case is applied element-wise and added up. The penalty function is designed so that the resulting cost function is smooth (twice differentiable).  However, if $a$ is violates the upper bound by $10\%$, a penalty of approximately $10^7$ is incurred--a high enough penalty so that if a feasible solution exists across all scenarios, it will be found.\\

The cost function $\mathrm{Cost}(\xv(t),\xv(t+1),t)$ is computed at each time step in the control period, but it requires state information from both $t$ and $t+1$ so it can incorporate generator ramping limits. The cost includes seven terms that penalize both economic cost of supplying generation and deviations of the system state outside of normal operational bounds.  The individual terms are:
\begin{itemize}
\setlength{\itemsep}{-2 pt}
\item[1] Generation costs
\[\GenCost(\pv_{\Go}(t))=\sum_{g \in \Go} c_{g1}({\pvg_{g}})^2+c_{g2}\pvg_{g}+c_{g3}.\]
\item[2] Generation limit penalties \[\Pen(\pvg(t),\pvgl,\pvgh).\]
\item[3] Ramping limit penalties \[\Pen\left(\pvgrl,\frac{\pvg(t+1)-\pvg(t)}{\tstep},\pvgrh\right).\]
\item[4] Power flow thermal limit penalties
    \[\sum_{\ije \in \E} \Pen(\pv_{\ije}(t),-\pvh_{\ije},\pvh).\]
\item[5] Frequency deviation penalties \[\Pen(\omega(t),-0.01,0.01)\]
\item[6] Integral frequency deviation penalties  \[\Pen(\Omega(t),-0.01,0.01).\]
\item[7] An integral deviation penalty on generation:  \[\Pen(\pvgi(T),0.95\Ilim,1.05\Ilim) \]
\end{itemize}
Cost 1 simply represents the financial cost of energy from different generators.  Costs 2-4 are normal power system constraints converted to costs using the penalty function defined above.  Cost 5 is an additional penalty designed to constrain the system frequency to within a 10 mHz band, and Cost 6 is designed to constrain the deviation of the integral of the frequency deviation so that the frequency is not allowed to be low or high for extended periods of time.  Finally, Cost 7 is designed to keep the total energy delivered by each controllable generator over the control period within a $\pm 5\%$ band around a $\pvgi(T)$ mimicking constraint the would occur in either a run-of-river hydro system or an energy storage device.

\subsection{Ensemble Optimal Control}
The evolution equations listed in \eqref{eq:Dyn} are functions of a given scenario $\scs$, therefore, we can think of the state  as a function of the scenario $\scs$ and the control parameters $\param=\{\paramp,\parami,\paramf,\pvgnot(t):0\leq t\leq T\}$: $\xv(\param,\scs,t)$. The overall optimization problem can then be written
\begin{align}\label{eq:OptProb}
&\min_{\param}  \sum_{\scs} \mathrm{Prob}(\scs)\left(\sum_{\tau=0}^{T-1} \mathrm{Cost}(\xv(\param,\scs,t),\xv(\param,\scs,t+1),t)\right) \nonumber\\
&\text{Subject to } \eqref{eq:Dyn}.
\end{align}
We optimize this objective using a standard numerical optimization algorithm (LBFGS \cite{minFunc}). The gradients of the objective function can be computed efficiently using a forward propagation algorithm that uses the chain rule to propagate gradients in time. This computation can be easily vectorized over all the scenarios, leading to significant speedup if run on a cluster or on GPUs.

\section{Numerical Results}
\label{sec:Numerics}

\begin{figure}
\begin{center}
\begin{subfigure}{.25\textwidth}
\includegraphics[width=\textwidth]{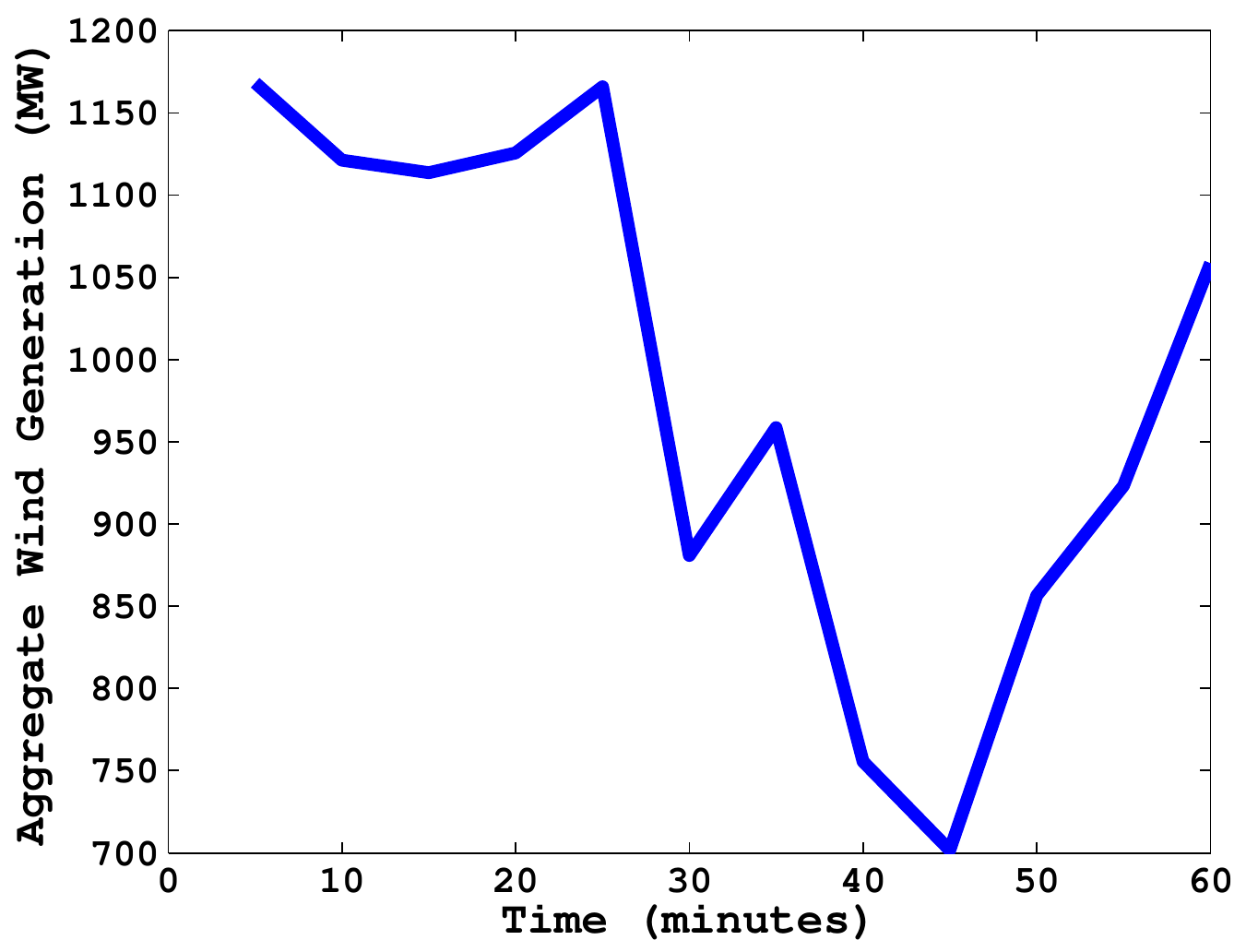}
\caption{Aggregate Wind Generation}
\label{fig:WindGen}
\end{subfigure}
\begin{subfigure}{.27\textwidth}
\includegraphics[width=\textwidth]{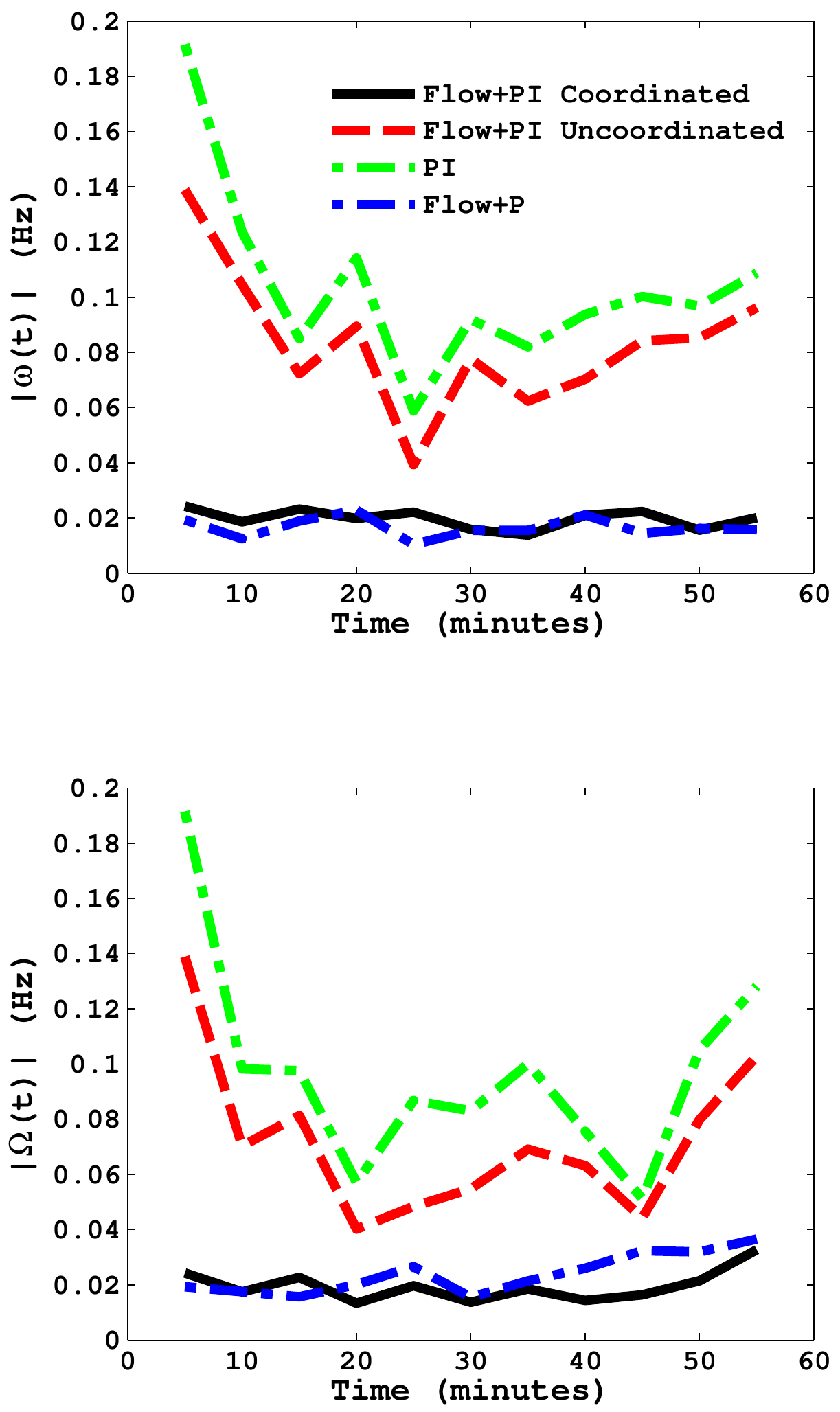}
\caption{Instantaneous and Integral Frequency Deviations}
\label{fig:Freq}
\end{subfigure}
\end{center}
\caption{Comparison of control schemes. a) Aggregate wind generation from a period with significant ramping events. b) {\it Worst-case} frequency deviations over the control period for 18 validation scenarios not used in the control design.       }
\end{figure}

\subsection{Description of Test System}\label{sec:test_system}

We test our algorithm using publicly available historical data for hydro and thermal generation, wind generation, and load from the Bonneville Power Administration (BPA) website\cite{BPA}.  We use a model of the BPA transmission system (shown in Fig.~\ref{fig:BPA}) that has 2209 buses and 2866 transmission lines. By identifying major hydroelectric stations on the transmission system and overlaying this onto a publicly available BPA wind site map \cite{BPA_wind_map}, we located the existing wind farms on the BPA transmission system (as of January 2010).  We located the meteorological stations where  BPA collects wind data \cite{BPA_met_map} in a similar manner. Using the same overlay, we used a simple incompressible air-flow model to infer hub height wind speeds at the wind farms.  The resulting wind speeds were passed through the power curve of a standard 1.5-MW GE Wind Turbine which was scaled to the wind farm nameplate capacity to estimate the power output $\pvr(t)$ (in MW) at each wind farm as a function of time. When we aggregate our wind farm-specific estimates of wind generation, we typically over estimate the BPA aggregate data by 20\%, which may caused by several factors including: spilling of wind by BPA, under performance of wind farms relative to single-turbine estimates, or shortcomings in our model of interpolating wind speeds.  BPA also provides aggregate load data\cite{BPA} that we divide among the nodes in the network according to population densities.  BPA also makes publicly available aggregate interchange flows \cite{BPA}, which we apportion to different tie lines in a similar manner.\\

  To test our control algorithm on difficult conditions, we select a control period of one hour from 10:35 AM to 11:35 AM on February 12, 2010, when the wind generation was ramping significantly (shown in Fig.~\ref{fig:WindGen}). We then create 26 scenarios (site-specific wind profiles) for this period by adding random time-varying Gaussian noise to the wind speeds at each meteorological station (from which we infer site-specific wind generation as outlined above).  We set the magnitude of the noise so as to match, on average, the aggregate wind generation hour-ahead forecast errors reported by BPA \cite{BPA_wind_forecast}. All the time series data used in our study was available at a 5-minute resolution.  \\

Unit commitment data is missing from our model, therefore, we assume that all hydro generators larger than 300 MW are online and are all participating in frequency regulation.  From inspection of the BPA historical generation data \cite{BPA}, we infer that the thermal generation dispatch is fixed over time.  In our model, we replicate this dispatch by dividing the total thermal generation among the online thermal generators (randomly chosen).

\subsection{Comparison of Various Control Schemes}

For difficult wind ramping conditions, we illustrate the value of feedback based on local flows by comparing four control schemes. We use P to designate proportional control (to frequency deviations $\omega(t)$) and I designates integral control (to integral frequency deviations $\Omega(t)$). The control schemes we consider using are:
\begin{itemize}
\item[1] \emph{PI:} Joint optimization of the time-varying dispatch $\pvgnot(t)$ and the local feedback parameters for $\omega(t)$ and $\Omega(t)$.
\item[2] \emph{Flow+PI Uncoordinated:} Time-varying dispatch $\pvgnot(t)$ plus feedback on $\omega(t),\Omega(t)$ and local flows $\pv_{\gi}$ at each generator. The optimization in 1 is performed first followed by a second optimization over the flow feedback parameters.
\item[3] \emph{Flow+PI  Coordinated:} Same as 2, but the optimization is performed jointly.
\item[4] \emph{Flow+P:} Same as 3, but without feedback on  $\Omega(t)$.
\end{itemize}

The experimental protocol is as follows. We setup each of the four optimization problems according to Eqs.~\ref{eq:OptProb} with the scenarios described in Section~\ref{sec:test_system} and determine a single set of feedback parameters for each of the four feedback schemes. We use 8 of the 26 created scenarios as input to the optimization algorithm. The remaining 18 unseen scenarios are reserved for validation of the control policy discovered by the optimization algorithm.  We note that all four control strategies are able to achieve similar generation costs while maintaining all the other constraints (line thermal capacities, ramping limits, and integral energy constraints), however, there are significant differences in the quality of the frequency regulation. Figure \ref{fig:Freq} shows the worst-case frequency deviations over the 18 validation scenarios. The frequency deviations are at an unacceptable level (.1-.2 Hz) when using just PI feedback (scheme 1).  If the flow feedback is included but optimized separately (scheme 2), there is little improvement.  However, the if the PI and flow feedback are coordinated via joint optimization (scheme 3), the frequency deviations are reduced to an acceptable level.  Interestingly, removing the feedback on the integral of the frequency deviations (scheme 4) does not impact the frequency deviations significantly relative to scheme 3.


\subsection{Discussion of the Results}

The distributed frequency control method we have presented benefits greatly from the incorporation of local power flows as demonstrated in Fig.~\ref{fig:Freq}.  There are several possible reasons for this improved performance.  First, power flows make the local generation-load imbalances visible to the generators so that the closest generators respond, effectively screening the more distant generators from the need to respond.  When compared to feedback based on frequency deviation, which is a global measure of the imbalance, feedback on local power flows confines imbalances to shorter spatial scales with a corresponding decrease in the time scale of the response.  An alternative explanation is that the optimization over the ensemble of possible futures in Eq.~\ref{eq:OptProb} is acting as a sort of machine learning that encodes correlations  between the wind prediction errors and the resulting local power flows into the flow feedback parameters.  When wind prediction error occurs, the change in power flows drives the feedback to nearly compensate for the error without a frequency deviation existing for any significant length of time. More numerical experiments are required to distinguish between these two (and other) possibilities.  In both of the possibilities discussed above, variations in the local power flows appear to be acting as ``pseudo-communication'' channels between the renewable and controllable generators.  Such a communication analogy may help explain why the independent optimizations in scheme 2 does not yield significant improvement in control performance.  The first optimization over frequency deviations may effectively washout the important local information in the power flows such that it is not available when optimizing over power flows.

\section{Conclusions and Future Work}
\label{sec:Future}

We introduced a control architecture based on off-line centralized optimization that can occur on a slow time scale coupled that sets the feedback parameters for fast distributed control of generation. The control scheme takes into account explicitly the variability in renewable generation using ensemble control. We showed that local feedback based on line flows and frequency deviations is sufficient to maintain all operational constraints and limit frequency deviations to an acceptable level even when the system is experiencing significant ramps in wind generation.  Our method exploits the hour-scale predictability of wind energy while using the off-line optimization to re-adjust control policies over longer timescales where wind predictability suffers. Our hybrid approach has the potential enable even higher levels time-intermittent renewable generation than presented here, and it can do so without real-time computation or communication. \\

These results are quite exciting and promising, however, they are preliminary and much work needs to be done to ensure the viability of this scheme in practice.
\begin{itemize}
\item Dynamical simulations are needed to check the dynamical stability of a grid with flow feedback. If these simulations show that the scheme is unstable, we believe that this can be rectified by appropriate exciter control at the generators to damp the fast electro-mechanical transients.

 \item   The scenario approach can be extended to include the (N-1) security criterion, so that the optimized  control strategy can deal with contingencies arising from the failure of a grid component. 

\item It is possible that flow feedback acts as a pseudo-communication channel between generators in the absence of a dedicated communication channel. It would be interesting to investigate this from an information theoretic point of view and investigate how much of information can be encoded in the flows.

\item We have used the simplest possible algorithmic approach by defining a smooth version of the optimization problem using penalty functions solving it using a generic LBFGS algorithm \cite{minFunc}. Second-order algorithms such as Stagewise Newton\cite{pantoja1983algorithms} or Differential Dynamic Programming (DDP)\cite{jacobson1968second} efficiently exploit the problem structure of deterministic optimal control problems. These can be leveraged in our ensemble control context by noting that when the feedback parameters $\parami,\paramp,\paramf$ are fixed, we have a deterministic optimal control problem in $\pvgnot(t)$ for each scenario. We have also been working on a Gauss-Newton algorithm for optimizing the fixed feedback$\parami,\paramp,\paramf$ efficiently. One can perform alternate minimization of $\pvgnot(t)$ and $\parami,\paramp,\paramf$ to get an efficient algorithm for optimizing both. Further, we note that when feedback does not include the integral term $\Omega(t)$, the ensemble control problem is a convex programming problem, and the global optimum can be found efficiently using specialized convex optimization techniques.

\item We plan to incorporate more accurate AC modeling of power flows taking advantage of most recent advances in analysis and algorithms related to optimizations of nonlinear power flows, e.g. \cite{12LL,BP-Energy}.

\item The integral energy constraint we introduced can also model energy storage, and our algorithm can easily be extended to incorporate distributed control of energy storage.
\end{itemize}

\section*{Acknowledgments}

The work at LANL was carried out under the auspices of the National Nuclear Security Administration of the U.S. Department of Energy at Los Alamos National Laboratory under Contract No. DE-AC52-06NA25396. SB and MC acknowledge partial support of NMC via NSF collaborative research grant ECCS-1128325 on ``Power Grid Spectroscopy''.

\bibliographystyle{unsrt}
\bibliography{DistributedAGC}

\end{document}